\def\year{2021}\relax
\documentclass[letterpaper]{article} 
\usepackage{aaai21}  
\usepackage{times}  
\usepackage{helvet} 
\usepackage{courier}  
\usepackage[hyphens]{url}  
\usepackage{graphicx} 
\usepackage{natbib}  
\usepackage{caption} 
\usepackage[dvipsnames]{xcolor}

\usepackage[super]{nth}
\usepackage{subcaption}
\usepackage{booktabs}
\usepackage{verbatim}
\usepackage[switch]{lineno}
\newcommand*\samethanks[1][\value{footnote}]{\footnotemark[#1]}

\title{Cough Against COVID: Evidence of COVID-19 Signature in Cough Sounds}

\author {
    Piyush Bagad\textsuperscript{\rm 1}\thanks{These authors contributed equally to this research}, Aman Dalmia\textsuperscript{\rm 1}\samethanks, Jigar Doshi\textsuperscript{\rm 1}\samethanks, Arsha Nagrani\textsuperscript{\rm 2}\thanks{Work done at Wadhwani AI as a Visiting Researcher}, \\
    Parag Bhamare\textsuperscript{\rm 1}, Amrita Mahale\textsuperscript{\rm 1}, Saurabh Rane\textsuperscript{\rm 1}, \\
    Neeraj Agarwal\textsuperscript{\rm 1}, Rahul Panicker\textsuperscript{\rm 1} \\
}
\affiliations {
    \textsuperscript{\rm 1} Wadhwani Institute for Artificial Intelligence \\
    \textsuperscript{\rm 2} VGG, Dept of Engineering Science, University of Oxford \\
}

\begin{document}

\maketitle

\begin{figure*}[!htp]
\centering
  \includegraphics[width=\linewidth]{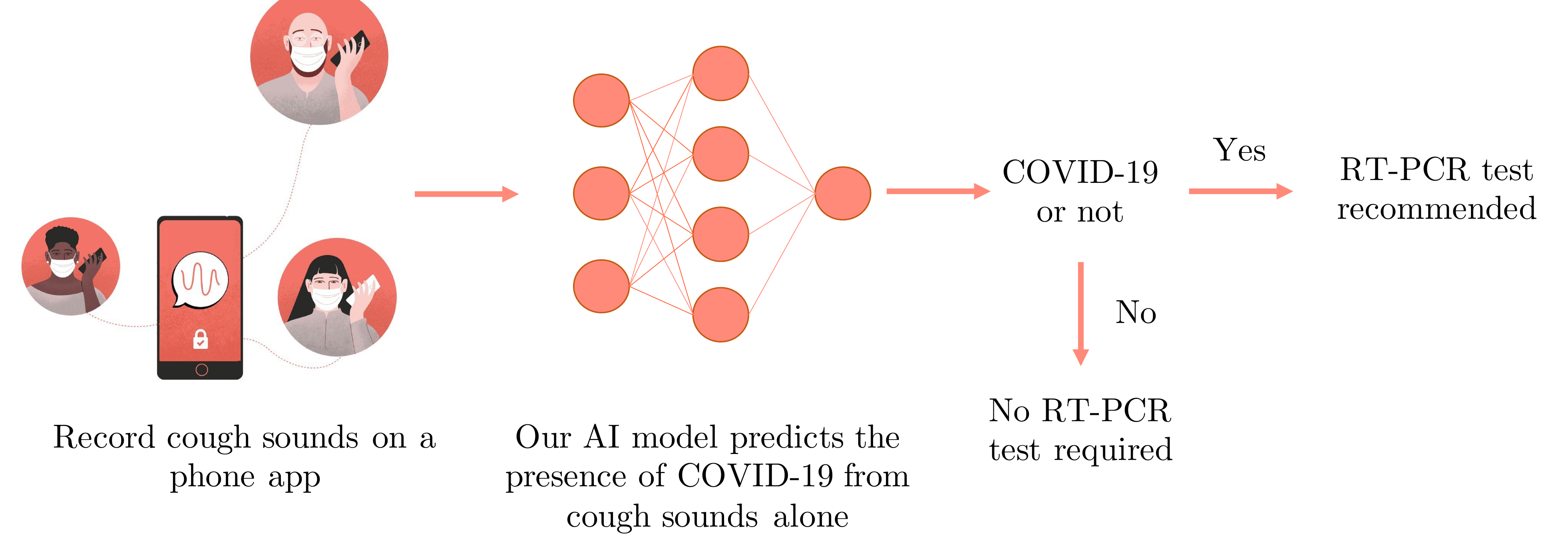}
  \caption{\textit{Cough Against Covid.} An overview of our non-invasive, AI-based pre-screening tool that determines COVID-19 status from solicited-cough sounds. With the AI model set to an operating point of high sensitivity, an individual is referred for gold standard RT-PCR test if they triage positive for risk of COVID-19. At 5\% disease prevalence, this triaging tool would increase the effective testing capacity by 43\%.}
  \label{fig:teaser}
\end{figure*}

\begin{abstract}
Testing capacity for COVID-19 remains a challenge globally due to the lack of adequate supplies, trained personnel, and sample-processing equipment. 
These problems are even more acute in rural and underdeveloped regions. 
We demonstrate that solicited-cough sounds collected over a phone, when analysed by our AI model, have statistically significant signal indicative of COVID-19 status (AUC 0.72, t-test, $p<0.01$, 95\% CI 0.61---0.83). 
This holds true for asymptomatic patients as well.
Towards this, we collect the largest known (to date) dataset of microbiologically confirmed COVID-19 cough sounds from 3,621 individuals. 
When used in a triaging step within an overall testing protocol, by enabling risk-stratification of individuals before confirmatory tests, our tool can increase the testing capacity of a healthcare system by 43\% at disease prevalence of 5\%, without additional supplies, trained personnel, or physical infrastructure.
\end{abstract}

\section{Introduction}
\label{sec:intro}

On \nth{11} March, 2020, the World Health Organisation (WHO) declared COVID-19 (also known as the coronavirus disease, caused by SARS-CoV2) a global pandemic. As of \nth{20} August, 2020, there were more than 22M confirmed cases of COVID-19 globally and over 788K deaths \cite{JHU_stats}. 
Additionally, COVID-19 is still active, with 267K new cases and 6,030 deaths per day world wide. As we eagerly await new drug and vaccine discoveries, a highly effective method to control the spread of the virus is frequent testing and quarantine at scale to reduce transmission \cite{Kucharski2020.04.23.20077024}. This has led to a desperate need for triaging and diagnostic solutions that can scale globally.

While the WHO has identified the key symptoms for COVID-19 -- fever, cough, and breathing difficulties, and recently, an expanded list \cite{WHO_QnA}, these symptoms are non-specific, and can deluge healthcare systems. Fever, the most common symptom, is indicative of a very wide variety of infections; combining it with a cough reduces the possible etiologies to acute respiratory infections (ARIs), which affect millions at any given time. Additionally, the majority of COVID-19 positive individuals show none of the above symptoms (asymptomatics) but they continue to be contagious \cite{WHO_report46, doi:10.7326/M20-3012, Daym1375}. To address this challenge, we present an AI-based triaging tool to increase the effective testing capacity of a given public health system. At the current model performance and at a prevalence of 5--30\%, our tool can increase testing capacity by 43--33\%. 

There have been various successful efforts using CT scans and X-rays to classify COVID-19 from other viral infections \cite{Wang2020COVIDNetAT, 2020arXiv200402060H, gozes2020rapid, He2020.04.13.20063941}. This suggests that COVID-19 affects the respiratory system in a characteristic way \cite{HUANG2020497, imran2020ai4covid} (see Section II (B) of \cite{imran2020ai4covid} for a detailed summary). The respiratory system is a key pathway for humans to both cough and produce voice – where air from the lungs passes through and is shaped by the airways, the mouth and nasal cavities. Respiratory diseases can affect the sound of someone’s breathing, coughing, and vocal quality – as most readers will be familiar with from having e.g. the common cold. Following this intuition we investigate whether there is a COVID-19 signature in solicited cough sounds and if it can be detected by an AI-model. 

The main contributions of this paper are as follows: (i) We demonstrate with statistical significance that solicited-cough sounds have a detectable COVID-19 signature;
(ii) Our modelling approach achieves a performance of 72\% AUC (area under the ROC curve) on held out subsections of our collected dataset;
(iii) We demonstrate with statistical significance that solicited-cough sound has a detectable COVID-19 signature among \textit{only asymptomatic} patients (Fig. \ref{fig:symptomatic}); (iv) We collect a large dataset of cough sounds paired with individual metadata and COVID-19 test results. To the best of our knowledge this is currently the largest cough dataset with verified ground truth labels from COVID-19 Reverse Transcription Polymerase Chain Reaction (RT-PCR) test results; and (v) Finally, we describe a triaging use case and demonstrate how our model can increase the testing capacity of the public health system by 43\%.


\section{Motivation and Related Work}
\label{sec:related}
Sound has long been used as an indicator for health. Skilled physicians often use stethoscopes to detect the presence of abnormalities by listening to sound from the heart or the lungs. Machine learning (ML), in particular, deep learning, has shown great promise in automated audio interpretation to screen for various diseases like asthma~\cite{oletic2016energy} and wheezing~\cite{li2017design} using sounds from smartphones and wearables. Open-source datasets like AudioSet~\cite{gemmeke2017audio} and Freesound Database~\cite{fonseca2018general} have further boosted research in this domain.



Automated reading of chest X-rays and CT scans~\cite{Wang2020COVIDNetAT, 2020arXiv200402060H, gozes2020rapid, He2020.04.13.20063941} have been widely studied along with typically collected healthcare data~\cite{Soltan2020.07.07.20148361} to screen for COVID-19. Respiratory sounds have also been explored for diagnosis (see~\cite{deshpande2020overview} for a nice overview). Some research has explored the use of digital stethoscope data from lung auscultation as a diagnostic signal for COVID-19~\cite{hui2020respiratory}. The use of human-generated audio as a biomarker offers enormous potential for early diagnosis, as well as for affordable and accessible solutions which could be rolled out at scale through commodity devices. 

Cough is a symptom of many respiratory infections. Triaging solely from cough sounds can be simple operationally and help reduce load on the healthcare system. ~\cite{saba2018techniques, botha2018detection} detect tuberculosis (TB) from cough sounds, while ~\cite{larson2012validation} track the recovery of TB patients using cough detection. A preliminary study on detecting COVID-19 from coughs uses a cohort of 48 COVID-19 tested patients versus other pathology coughs to train a combination of deep and shallow models~\cite{imran2020ai4covid}. Other valuable work in this domain investigates a similar problem~\cite{brown2020exploring}, wherein a binary COVID-19 prediction model is trained on a dataset of crowdsourced, unconstrained worldwide coughs and breathing sounds.  In~\cite{han2020early} speech recordings from COVID-19 hospital patients are analyzed to automatically categorize the health state of patients. A crowdsourced dataset~\cite{sharma2020coswara} of cough, breathing and voice sounds was also recently released to enable sound as a medium for point-of-care diagnosis for COVID-19.

Apart from ~\cite{imran2020ai4covid} and ~\cite{brown2020exploring}, none of the previous efforts actually detect COVID-19 from cough sounds alone. ~\cite{imran2020ai4covid} covers only 48 COVID-19 tested patients, while our dataset consists of 3,621 individuals with 2,001 COVID-19 tested positives. The dataset used in ~\cite{brown2020exploring} was entirely crowdsourced with the COVID-19 status being self-reported, whereas our dataset consists of labels directly received from healthcare authorities. Further, we show that COVID-19 can be detected from the cough sounds of \textit{asymptomatic} patients as well. Unlike previous works, we also demonstrate how label smoothing can help tackle the inherent label noise due to the sensitivity of the RT-PCR test and improve model calibration.

\section{Data}
\label{sec:data}
In this section we outline our data collection pipeline as well as the demographics and properties of the gathered data. We further describe the subset of the data used for the analysis in this paper.

We note here that we use two types of data in this work. First, we describe data collected from testing facilities and isolation wards for COVID-19 in various states of India, constituting the largest dataset of tested COVID-19 cough sounds (to the best of our knowledge). Next, we mention several open-source cough datasets that we use for pretraining our deep networks.

\begin{figure*}
    \centering
    \includegraphics[width=\linewidth]{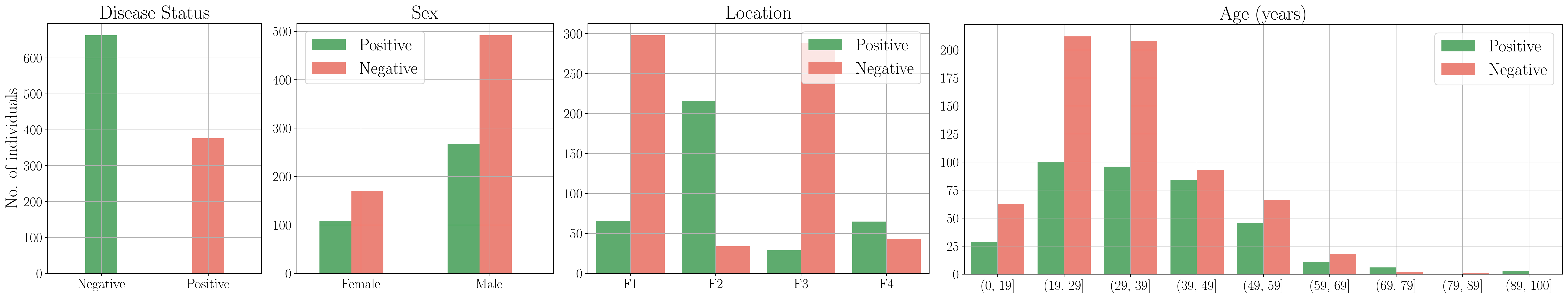}
    \caption{\textit{Dataset demographics.} From left to right -- distribution of the number of individuals based on COVID-19 test result, sex, location and age.}
    \label{fig:data_demo}
\end{figure*}



\subsection{COVID-19 cough dataset} 
\subsubsection{Data collection}
We create a dataset of cough sounds from COVID-19 tested individuals from numerous testing facilities and isolation wards throughout India (collection is ongoing). Testing facilities provide data for both positively and negatively tested individuals, whereas isolation wards are meant only for those who have already tested positive. For isolation wards, we only consider individuals within the first 10 days after an initial positive result through RT-PCR. Our eligibility criteria also requires that individuals should be independently mobile and be able to provide cough samples comfortably. The data collector is required to wear a PPE kit prior to initiating conversation, and maintain a distance of 5 feet at all times from the participant. The participant is required to wear a triple layer mask and provide written consent. For minors, consent is obtained from a legally acceptable representative. Our data collection and study have been approved by a number of local and regional ethics committees
\footnote{The names of the precise committees have been omitted to preserve anonymity, and will be added to any future versions.}.
\begin{figure}
     \centering
     \includegraphics[width=\linewidth]{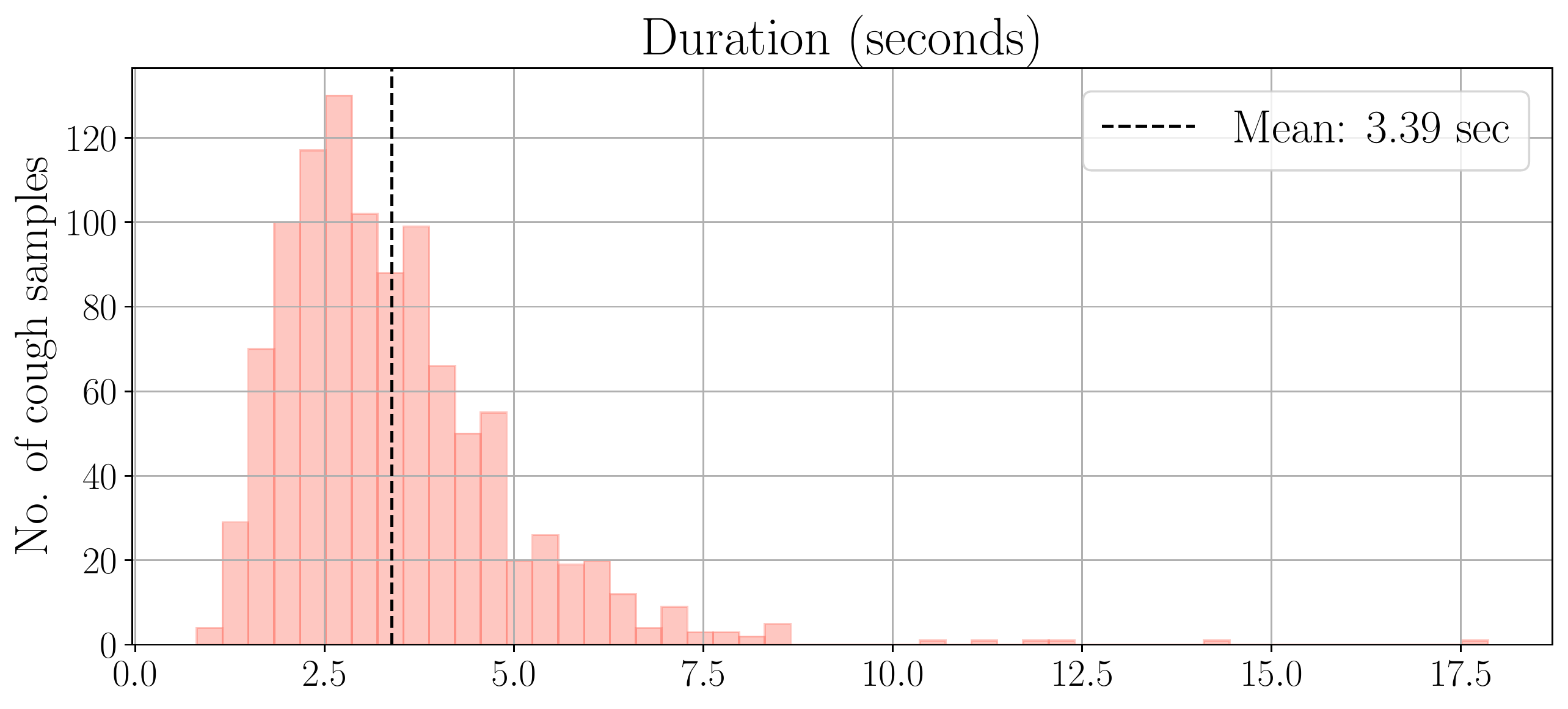}
    \caption{\textit{Duration statistics.} Distribution of the duration of the cough audio recordings.}
    \label{fig:data_facility_duration}
\end{figure}

For each individual, our data collection procedure consists of the following 3 key stages: 
\begin{enumerate}
    \item \textbf{Subject enrollment:} In the first stage, subjects are enrolled with metadata such as demographic information (including self-reported age and sex), the presence of symptoms such as dyspnea (shortness of breath), cough and fever, recent travel history, contact with known COVID-19 positive individuals, body temperature, and any comorbidities or habits such as smoking that might render them more vulnerable. 
    \item \textbf{Cough-sound recording:} Since cough is an aerosol generating procedure, recordings are collected in a designated space which is frequently disinfected as per facility protocol. For each individual, we collect 3 separately recorded audio samples of the individual coughing, an audio recording of the individual reciting the numbers from one to ten and a single recording of the individual breathing deeply. Note here that these are non-spontaneous coughs, i.e. the individual is asked to cough into the microphone in each case, even if they do not naturally have a cough as a symptom. 
    \item \textbf{Testing:} RT-PCR test results are obtained from the respective facility's authorized nodal officers. 
\end{enumerate}

For each stage, we utilise a separate application interface. Screenshots for the apps and further details are provided in suppl. material. We note here that the COVID-19 test result is \textit{not} known at the time of audio cough recording -- minimising collection bias, and that all data collection is performed in environments in which potential solutions may actually be used.

\begin{figure}[t]
    \centering
    \includegraphics[width=\linewidth]{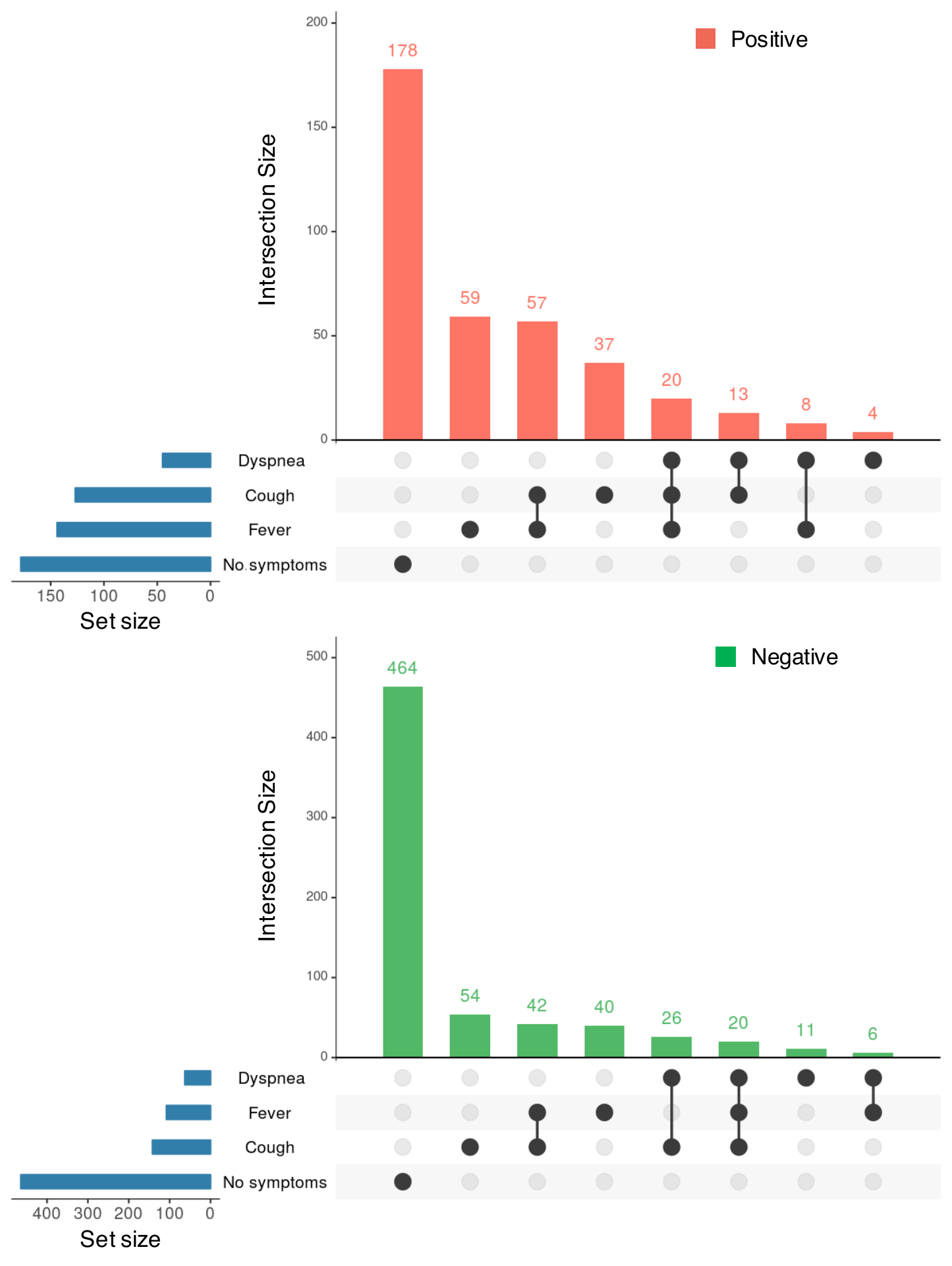}
    \caption{\textit{Symptom co-occurrence statistics.} We show statistics for individuals with an RT-PCR positive (top) and negative (bottom) test for the following symptoms: dyspnea (shortness of breath), cough and fever.}
    \label{fig:symptoms}
\end{figure}

\subsubsection{Dataset}
As of August 16th, 2020 our data collection efforts have yielded a dataset of 3,621 individuals, of which 2,001 have tested positive. In this paper we focus on a curated set of the collected data (until 20 July, 2020). We also restrict our models to use only the cough sounds (and not the voice or breathing samples). Henceforth, all results and statistics will be reported on this data used in our analysis after filtering and manual verification (details of which are provided in the suppl. material). 
Our curated dataset consists of 3,117 cough sounds from 1,039 individuals. We aim to release some or all of the data publicly to the research community. Figures \ref{fig:data_demo} and \ref{fig:data_facility_duration} show distribution statistics of the data. Out of 1,039 individuals, 376 have a positive RT-PCR test result (Fig. \ref{fig:data_demo}, left) and the sex breakdown is 760 male and 279 female. (Fig. \ref{fig:data_demo}, center-left). (Fig. \ref{fig:data_demo}, center-right) highlights the distribution by the facility from which the data was collected (we use data from 4 facilities, F1-F4). (Fig. \ref{fig:data_demo}, right) shows the age distribution, which is skewed towards middle-aged individuals (between 20-40 years of age), while Fig. \ref{fig:data_facility_duration} shows the distribution of the lengths of our cough samples.
Fig \ref{fig:symptoms} shows the distribution of symptoms recorded for dyspnea, cough and fever. Interestingly, note that most individuals are asymptomatic. In our dataset, the most common single symptom among COVID-19 positive individuals is fever while that among negatives is cough, followed by an intersection of cough and fever. 

\begin{figure*}[!ht]
\centering
  \includegraphics[width=0.8\textwidth]{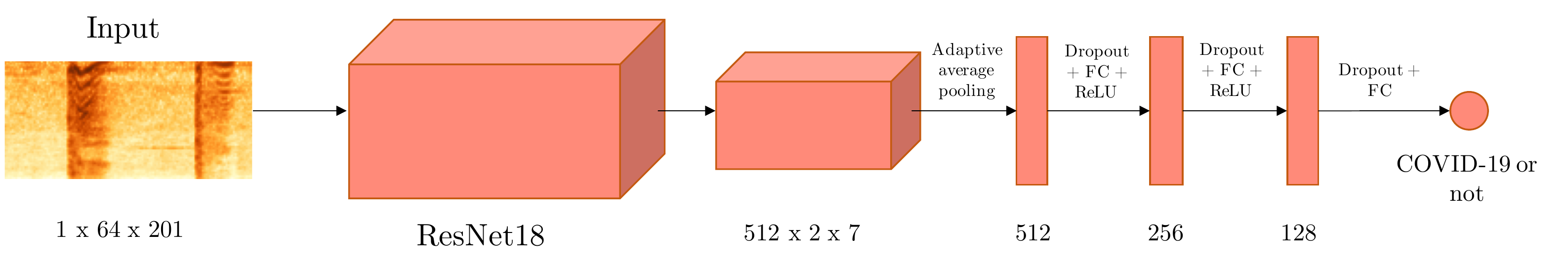}
  \caption{\textit{Network architecture.} An input cough spectrogram goes through a deep CNN to predict the presence of COVID-19.}
  \label{fig:arch}
\end{figure*}
\subsection{Open-source non-COVID cough datasets}
\label{sec:pretrained-datasets}
In the absence of explicit feature engineering, deep Convolutional Neural Networks (CNNs) are data hungry – relying on thousands of manually annotated training examples. Given the challenges of training deep CNNs from scratch on small datasets, we collect a larger dataset of cough samples from various public datasets \cite{fonseca2018general,al2020flusense,sharma2020coswara} which we use to pretrain our model. In total we obtain 31,909 sounds segments, of which 27,116 are non-cough respiratory sounds (wheezes, crackles or breathing) or human speech, and 4,793 are cough sounds. The various data sources and their statistics are as follows: \\
\noindent\textbf{1. FreeSound Database 2018~\cite{fonseca2018general}:} This is an audio dataset consisting of a total of 11,073 audio files annotated with 41 possible labels, of which 273 samples are labelled as `cough'. We believe the cough sounds correspond to COVID-19 negative individuals as these sounds were recorded well before the COVID-19 pandemic. \\
\noindent\textbf{2. Flusense~\cite{al2020flusense}:} This is a subset of Google's Audioset dataset~\cite{gemmeke2017audio}, consisting of numerous respiratory sounds.~\footnote{Including speech, coughs, sneezes, sniffles, silence, breathing, gasps, throat-clearing, vomit, hiccups, burps, snores, and wheezes.} We use 11,687 audio segments of which 2,486 are coughs. \\
\noindent\textbf{3.  Coswara~\cite{sharma2020coswara}:} This is a curated dataset of coughs collected via worldwide crowd sourcing using a website application\footnote{\url{https://coswara.iisc.ac.in/}}. The dataset contains samples from 570 individuals, with 9 voice samples for each individual, including breathing sounds (fast and slow), cough sounds (heavy and shallow), vowel sounds, and counting (fast and slow). In total the dataset consists of 2,034 cough samples and 7,115 non-cough samples. We are unaware of the COVID-19 status of the coughs in this dataset as it was collected after the pandemic broke out.

\section{Method}
\label{sec:method}
Inspired by the recent success of CNNs applied to audio inputs~\cite{hershey2016cnn}, we develop an end-to-end CNN-based framework that ingests audio samples and directly predicts a binary classification label indicating the probability of the presence of COVID-19. In the following sections, we outline details of the input, model architecture, training strategies employed and inference. 

\subsection{Input}
During training we randomly sample a 2-second segment of audio from the entire cough segment. We use short-term magnitude spectrograms as input to our CNN model. All audio is first converted to single-channel, 16-bit streams at a 16kHz sampling rate for consistency. Spectrograms are then generated in a sliding window fashion using a hamming window of width 32ms and hop 10ms with a 512-point FFT. This gives spectrograms of size 257 x 201 for 2 seconds of audio.
The resulting spectrogram is integrated into 64 mel-spaced frequency bins with minimum frequency 125Hz and maximum frequency 7.5KHz, and the magnitude of each bin is log transformed.
This gives log-melspectrogram patches of 64 x 201 bins that form the input to all classifiers.
Finally, the input is rescaled by the largest magnitude over the training set to bring the inputs between -1 and 1.

\subsection{CNN architecture}
An overview of our CNN architecture can be seen in Fig.~\ref{fig:arch}. As a backbone for our CNN model we use the popular ResNet-18 model consisting of residual convolution layers \cite{he2016deep}, followed by adaptive pooling layer in both the time and frequency dimensions. Finally, the output is passed through 2 linear layers and then a final predictive layer with 2 neurons and a softmax activation function, which is used to predict whether the input cough sample has COVID-19. Dropout~\cite{srivastava2014dropout} and the ReLU activation function are used after all linear layers. 


\subsection{Training strategies}
\subsubsection{Augmentation}
Given the medium size of our dataset, we adopt the standard practise of data augmentation, applying transformations to our data to boost performance and increase robustness. We perform data augmentation online, i.e.\ transformations are applied randomly to segments during training. We perform two types of augmentation: (1) the addition of external background environmental sounds from the ESC-50 dataset~\cite{piczak2015esc}, and (2) time and frequency masking of the spectrogram input~\cite{park2019specaugment}. 
ESC-50~\cite{piczak2015esc} consists of 2,000 environmental audio recordings from 50 environmental classes. At train time, we randomly select a single noise sample and modulate the amplitude by a random factor between 0.4 and 0.75, before adding it to the input cough spectrogram.  

\subsubsection{Pre-training}
Our model architecture is first pretrained on the open source cough datasets outlined in Sec. \ref{sec:pretrained-datasets}.
We partition the data into train and validation (the validation set consists of 648 cough and 2882 non-cough sounds), and train our model to simply predict the presence of a cough or not (cough detection). Note that this is a proxy task and we use this simply to pretrain our model and learn a good initialisation of weights. 

We first initialise the ResNet-18 backbone with weights obtained from pretraining on ImageNet (the additional linear layers after are initialised randomly). Given the highly unbalanced nature of the pretraining data, we upsample the minority class to ensure that each batch has the equal number of cough and non-cough samples. AdamW~\cite{loshchilov2017decoupled} is used as the optimizer with a learning rate of 1e-5 and weight decay 1e-4. The model is trained for 200 epochs and on the proxy cough vs. non-cough task, we achieve an AUC of 0.98 on the validation set. 

\subsubsection{Label smoothing}
For our final task of COVID-19 classification, we note here that the ground truth labels come solely from the RT-PCR test for COVID-19. Even though this test is widely used, it is known to make mistakes, i.e. it is estimated to have a sensitivity of almost 70\% at a specificity of 95\%~\cite{Watsonm1808}. Hence it is possible that a number of cough samples may have the wrong label, and penalising our model for making mistakes on these samples can harm training. Hence we apply a standard label smoothing technique~\cite{mller2019does} during training for each instance. Label smoothing is also known to improve model calibration~\cite{mller2019does}. Results are provided in Sec. \ref{sec:smoothing}.

 

\subsubsection{Implementation details}
For cough classification, we use the pretrained weights from the cough non-cough pretraining task to initialize the model. SGD is used as the optimizer, with an initial learning rate of 0.001 and a decay of 0.95 after every 10 epochs. We use a batch size of 32 and train for a total of 110 epochs. Label smoothing is randomly applied between 0.1 and 0.3. Our model is implemented in PyTorch~\cite{paszke2019pytorch} (version 1.6.0) and trained using a single Tesla K80 GPU on the Linux operating system. The same seed has been set for all our experiments (more details can be found in suppl. material). We used Weights \& Biases~\cite{wandb} (version 0.9.1) for experiment tracking and visualisation.

\subsection{Inference}
Every cough sample is divided into 2-second segments using a sliding window with a hop length of 500ms. We take the median over the \textit{softmax} outputs for all the segments to obtain the prediction for a single sample. We pad inputs less than 2 seconds with zeros. A comparison of different aggregation methods have been provided in suppl. material.

\subsubsection{Individual-level aggregation}
For each individual in the dataset, we have three cough samples. We consider the \textit{max} of the  predicted probabilities of the three cough samples to obtain the prediction for a single individual. All performance metrics have been reported at the individual level. 


\section{Experimental Evaluation}
\subsection{Tasks} 
Although we train our model on the entire dataset once, we focus on three clinically meaningful evaluations: 

\begin{itemize}
    \item \textbf{Task 1:} Distinguish individuals tested \textit{positive} from individuals tested \textit{negative} for COVID-19.
    
    \item \textbf{Task 2:} Distinguish individuals tested \textit{positive}, from individuals tested \textit{negative} for COVID-19, specifically for individuals that do \textit{not report cough as a symptom}. We refer to this set as Asymptomatic (no C). 
    
    \item \textbf{Task 3:} Distinguish individuals tested \textit{positive}, from individuals tested \textit{negative} for COVID-19, specifically for individuals that do \textit{not report cough, fever or breathlessness as a symptom}. We refer to this set as Asymptomatic (no C/F/D).

\end{itemize}

The number of cough samples in the validation set for each task are provided in Table \ref{tbl:tasks-dataset}. Fig. \ref{fig:symptomatic} shows the comparison in performance across the three tasks.
\begin{table}[!h]
  \centering
    \begin{tabular}{l|l|l}
    \hline
        Task & Positive & Negative    \\ \hline
        (1) &  87-102 & 108-117 \\ \hline
        (2) & 57-75 & 78-105  \\ \hline
        (3)  & 45-66 & 69-93 \\ \hline
   \end{tabular}
  \caption{\textit{Dataset statistics per task.} Number of cough samples in the validation set for each task. Since we perform 5-fold validation, we show the range from min-max. Note that the precise number of samples varies across folds as we select 10\% of the total dataset but ensure that the validation set is balanced per facility. Note that each individual has three cough samples.}
  \label{tbl:tasks-dataset}
\end{table}

\subsection{Triple-stratified cross-validation} 
In order to create a fair evaluation, we (1) create training and validation sets from disjoint individuals, (2) we balance the number of positive and negatives obtained from each facility in the validation set, to ensure that we are not measuring a facility specific bias, and (3) we upsample the minority class samples per facility in the train set (facility-wise class distribution has been shown in Fig. \ref{fig:data_demo}). 
We split our dataset into train and validation sets of approximately 90\%:10\% ratio, and following standard practise for ML methods on small datasets,  perform 5-fold cross-validation.

\subsection{Evaluation metrics}
\label{sec:evaluation}
We report several standard evaluation metrics such as the Receiver Operating Characteristic - Area Under Curve (ROC-AUC), Specificity (1 - False Positive Rate (FPR)), and Sensitivity (also known as True Positive Rate (TPR)). Since this solution is  meant to be used as a triaging tool, high sensitivity is important. Hence, we report the best specificity at 90\% sensitivity. We report mean and standard deviation across all 5 cross-validation folds. For fairness, all hyperparameters are set on the first fold and applied, as is, to other folds, including epoch selection.

\subsection{Comparison to shallow baselines}\label{sec:classical}
We also compare our CNN-based model to shallow classifiers using hand-crafted audio features. We experiment with the following classifiers: (1) Logistic Regression (LR), (2) Gradient Boosting Trees (3) Extreme Gradient Boosting (XGBoost) and (4) Support Vector Machines (SVMs). As input to the classifiers, we use a range of features such as the tempo, RMS energy and MFCCs (see Sec. 4.1 from~\cite{brown2020exploring} for an exhaustive list of the features used.) 
For all methods, we follow the preprocessing design choices adopted by~\cite{brown2020exploring}. We optimize the hyperparameters following the same procedure outlined in \ref{sec:evaluation}. 

\subsection{Stacked ensemble}
We ensemble the individual-level predictions from ResNet-18 (both with and without label smoothing) and the XGBoost classifier (described in detail in Sec. \ref{sec:classical}) using Stacked Regression~\cite{van2007super}. The stacked regressor is a XGBoost classifier using the predicted probabilities from each of the above models as features. The hyperparameters for the regressor are mentioned in the suppl. material. We report performance with and without the ensemble (Fig. \ref{fig:ensemble-comp}).

\subsection{Ablation analysis}
We also quantify the effect of several aspects of our training pipeline, notably - pretraining, label smoothing and the length of the input segment. We experiment with two segment lengths - 1 second and 2 seconds. For the model trained on 1-second input segments, we perform hyperparameter tuning again. Results for all ablation analysis are provided in Sec. \ref{sec:Discussion}.

\section{Discussion} 
\label{sec:Discussion}
\begin{figure}[!ht]
    \centering
     \begin{subfigure}[b]{0.5\linewidth}
         \centering
         \includegraphics[width=\textwidth]{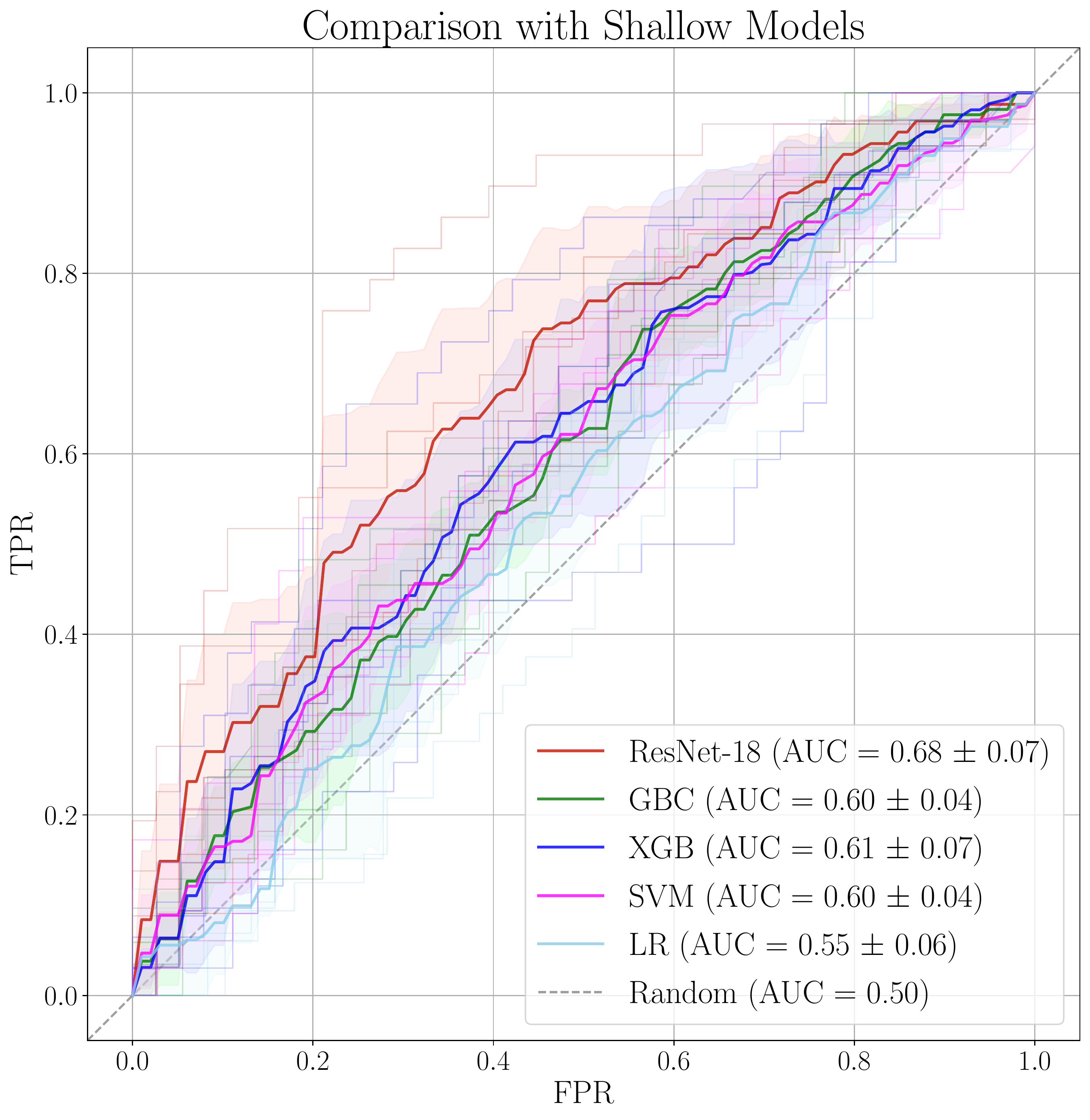}
         \caption{Shallow models vs CNN}
         \label{fig:model-comp}
     \end{subfigure}%
     \begin{subfigure}[b]{0.5\linewidth}
         \centering
         \includegraphics[width=\textwidth]{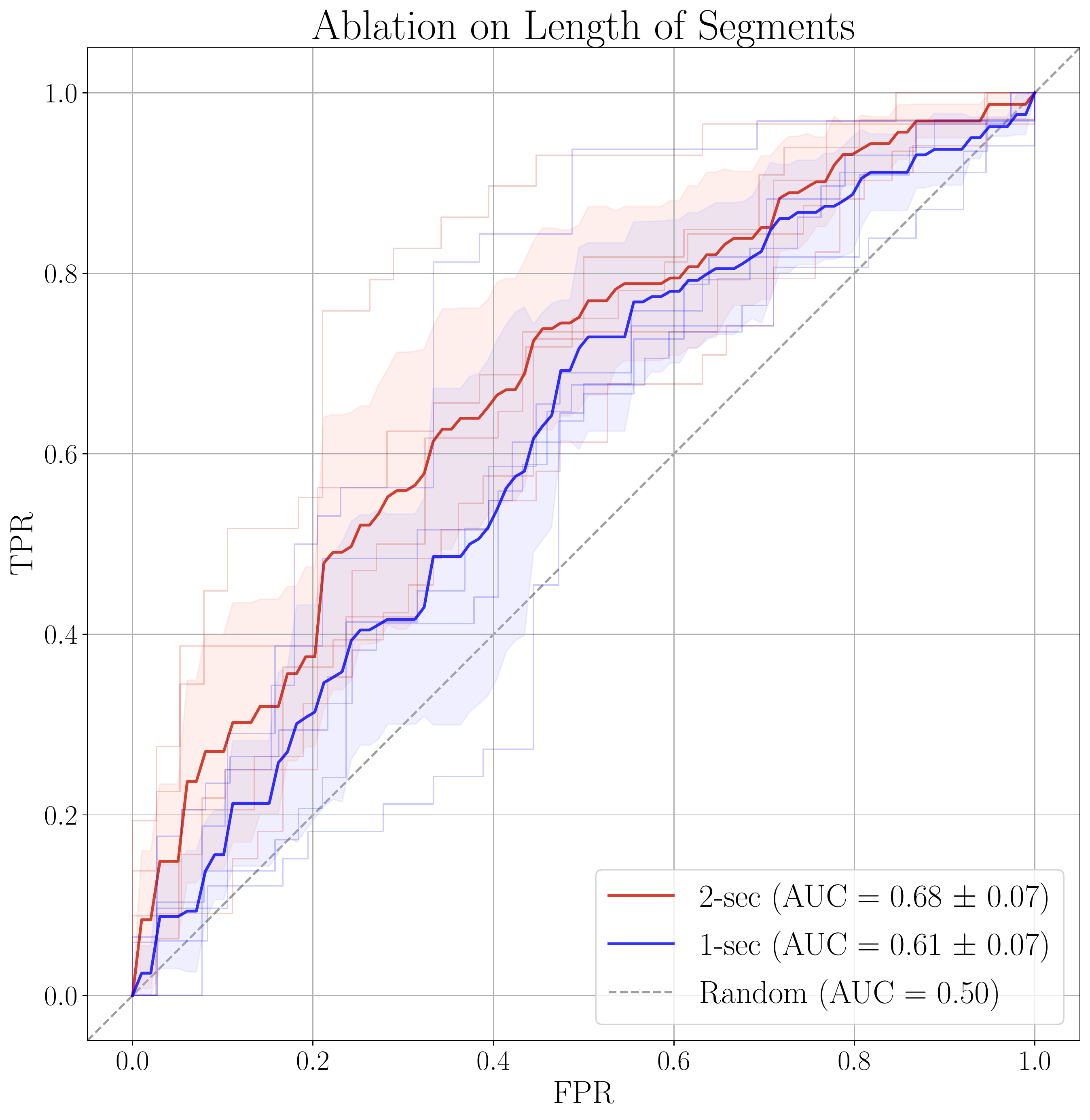}
         \caption{1-sec vs 2-sec segments}
         \label{fig:segments}
     \end{subfigure}
  \caption{\textit{Ablation results.} Comparison of ROC curves across (a) different model families - ResNet-18 outperforms other shallow baselines; (b) different segment lengths. 2-second is found to be the optimal segment length}
  \label{fig:ablation}
\end{figure}

Fig. \ref{fig:model-comp} shows that the CNN-based model outperforms all shallow models by atleast 7\% in terms of AUC. 
We also perform a statistical significance analysis of the results of our model. We conduct a $t$-test with the Null Hypothesis that there is no COVID-19 signature in the cough sounds and the results were found to be statistically significant, $p < 1e-3$, $95\%$ confidence interval (CI) $0.61$---$0.83$.

\subsection{Effect of ensembling}
\begin{figure}[!ht]
    \centering
    \begin{subfigure}[b]{0.5\linewidth}
         \centering
         \includegraphics[width=\textwidth]{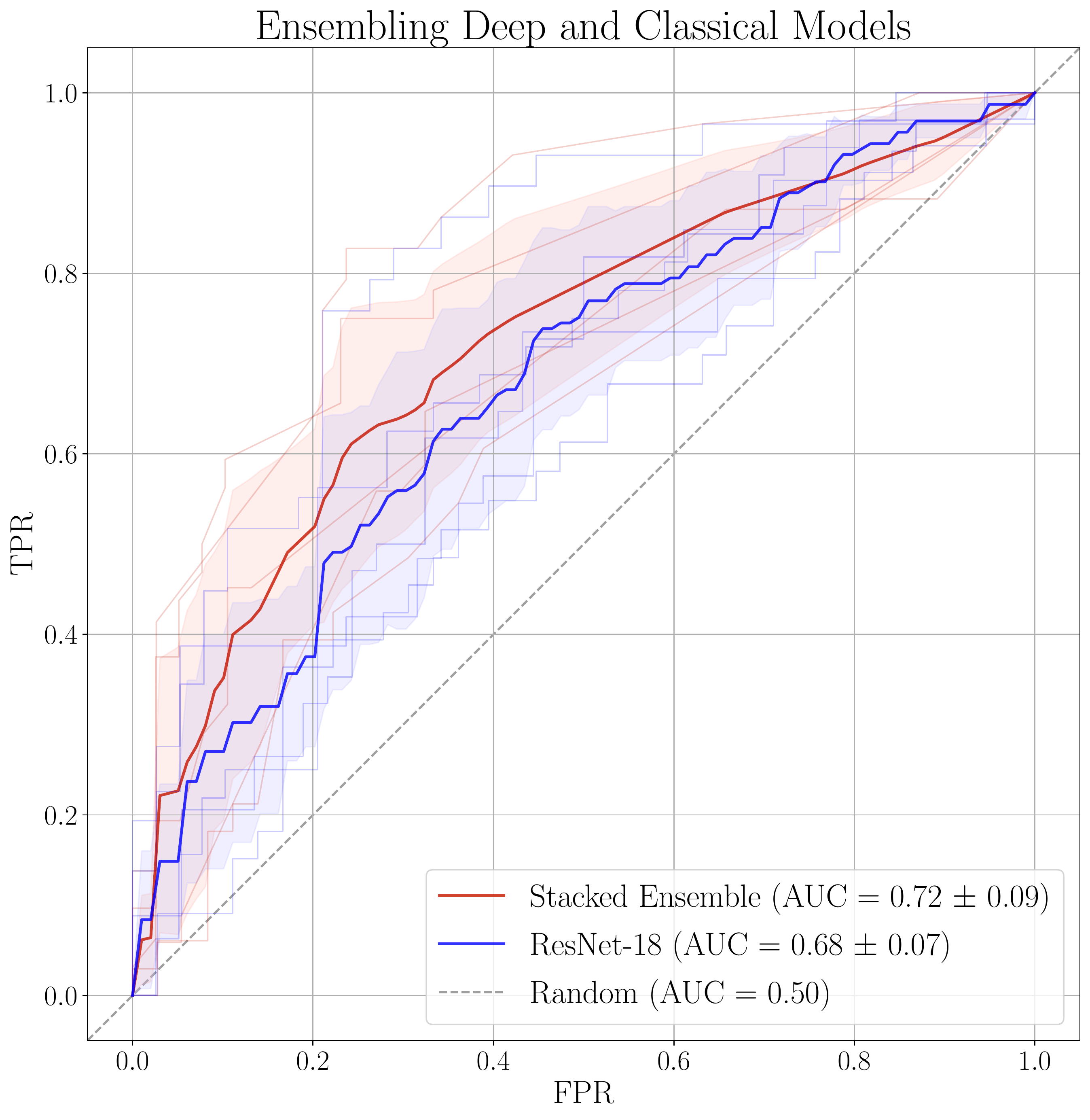}
         \caption{Ensembling}
         \label{fig:ensemble-comp}
     \end{subfigure}%
     \begin{subfigure}[b]{0.5\linewidth}
         \centering
         \includegraphics[width=\textwidth]{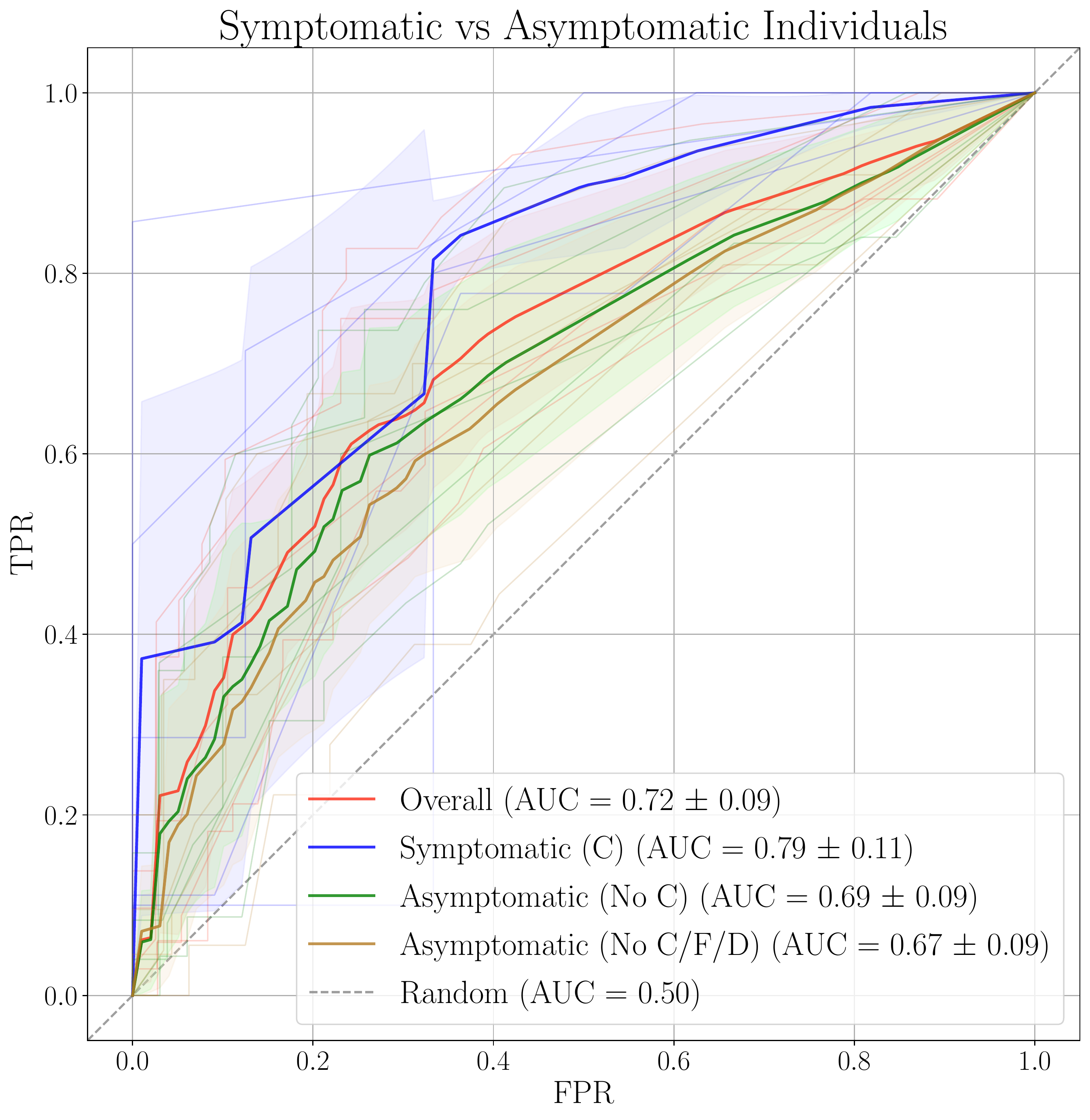}
         \caption{Asymptomatic individuals}
         \label{fig:symptomatic}
     \end{subfigure}
  \caption{\textit{Classification results.} Comparison of ROC curves (a) for our best model obtained by ensembling a shallow model with the deep model. (b) for symptomatic and asymptomatic individuals: C - cough, F - fever, D - dyspnea (shortness of breath). Our model is able to identify COVID-19 from the cough sounds of asymptomatic individuals as well.}
  \label{fig:results}
\end{figure}

It is widely known that ensembling diverse models can improve performance, even if some models perform worse than others individually~\cite{sagi2018ensemble}. Fig. \ref{fig:ensemble-comp} empirically validates this for our task by showing that ensembling the deep and shallow models improves performance compared to any of the individual models. This also indicates that there is further room for performance improvement through better ensembling techniques and using more diverse models.

\subsection{Effect of label smoothing} \label{sec:smoothing}
The effect of applying label smoothing has been reported in Table \ref{tbl:label-smoothing}. Besides improving AUC, label smoothing also improves the specificity at 90\% sensitivity. This shows that at the required operating point (threshold on the softmax scores) for a triaging tool, the model is able to classify better with smoothened labels. This suggests that explicitly dealing with label noise can improve performance. We also empirically verify that label smoothing improves model calibration \cite{mller2019does} as it drives the optimal threshold for COVID-19 classification much closer to 0.5.
\begin{table}[!h]
\centering
    \begin{tabular}{l|l|l|l}
    \hline
        Model & AUC          & Specificity & Threshold    \\ \hline
        with LS & 0.68 $\pm$ 0.07 & 0.31 $\pm$ 0.13                      & 0.422 $\pm$ 0.062 \\ \hline
        no LS   & 0.65 $\pm$ 0.08 & 0.27 $\pm$ 0.11                      & 0.002 $\pm$ 0.002 \\ \hline

\end{tabular}
  \caption{\textit{Effect of label smoothing.} Label smoothing improves specificity at 90\% sensitivity and model calibration.}
  \label{tbl:label-smoothing}
\end{table}

\subsection{Effect of pre-training}
Table \ref{tbl:pretraining} shows the utility of using pretrained weights. Pretraining improves the mean AUC by 17\%, showing it's importance in dealing with small or medium sized datasets like ours. 
\begin{table}[!h]
\centering
    \begin{tabular}{l|l}
    \hline
        Model & AUC    \\ \hline
        with pretraining & 0.68 $\pm$ 0.07 \\ \hline
        no pretraining   & 0.51 $\pm$ 0.07 \\ \hline
\end{tabular}
  \caption{\textit{Effect of pre-training.} Pre-training greatly improves model performance.}
  \label{tbl:pretraining}
\end{table}

\subsection{Optimal segment length}
Fig. \ref{fig:segments} indicates that using segments of 2-seconds performs better than 1-second segments. We suspect that this happens because our dataset contains several samples with silence at the start and the end, increasing the probability of noisy labels being assigned to random crops during training. 

\subsection{Asymptomatic individuals}
Fig. \ref{fig:symptomatic} shows the performance for asymptomatics. We see that while our model performs significantly better for symptomatic individuals, performance for asymptomatic individuals is still far above random.  A $t$-test was conducted with the Null Hypothesis that there is no COVID-19 signature in the cough sounds of asymptomatic patients and the results were found to be statistically significant, $p < 1e-2$.

\subsection{Performance across sex and location}
While we note that our dataset contains more males than females, there is no obvious bias in COVID-19 test results (Fig. \ref{fig:data_demo}), and performance is similar for both male ($0.71 \pm 0.11$) and female ($0.72 \pm 0.11$) individuals. 



Samples collected from different \textit{locations} can have different label distributions. For example, testing facilities (F1, F3 and F4) tend to have predominantly COVID-19 negatives while isolation wards (F2) tends to contain COVID-19 positives (Fig. \ref{fig:data_demo}). Naively training a classifier on this combined dataset would lead to significantly inflated performance because it could simply learn a \textit{location} classifier instead of a COVID-19 cough classifier. This is a known phenomenon in deep learning and medical imaging ~\cite{badgeley2019deep} ~\cite{wachinger2019quantifying}. To address this issue, we carefully constructed our validation set to contain only testing facilities with equal number of positive and negative samples per location. Future work will explore algorithmic mitigation by applying techniques such as ~\cite{zhang2018mitigating}.




\section{Use Case: COVID-19 Triaging Tool}
\label{sec:triaging tool}

In India alone, as of the 21st of August, 2020, there have been over 33M COVID19 RT-PCR tests performed \cite{ICMR_TestsInIndia}. While the current testing capacity is ~800k/day, the test positivity rate (TPR) has been increasing at a steady pace, indicating that there is an urgent need for testing to be ramped up even further. The ability to ramp up tests, however, is significantly hindered by the limited supply of testing kits and other operational requirements such as trained staff and lab infrastructure. This has led to an increased urgency for accurate, quick and non-invasive triaging, where individuals most likely to be determined positive for COVID19 are tested as a priority.

To address this, we propose a triaging tool that could be used by both individuals and health care officials. We pick the threshold of the model such that we have a high sensitivity of 90\% which is desirable for a triaging tool. At this sensitivity our best model has a specificity of 31\%. As shown in Fig. \ref{fig:teaser}, such a model can be used to reliably detect \textit{COVID-19 negative individuals} while we refer the positives for a confirmatory RT-PCR test. In this way, we increase the testing capacity by 43\% (a 1.43x lift) when we assume a disease prevalence of 5\%. In Table \ref{table:lift}, we also show the relative gains at different prevalence levels. Precise calculations can be found in the suppl. material. 

\begin{table}[!ht]
\centering
    \begin{tabular}{l|l}
    \hline
        Prevalence & Testing Capacity  \\ \hline
        1\% & +44\% \\ \hline
        5\%   & +43\% \\ \hline
        10\%   & +41\% \\ \hline
        30\%   & +33\% \\ \hline
\end{tabular}
  \caption{\textit{Utility of our triaging tool.} We show the increase in the effective testing capacity of a system at different disease prevalence levels.}
\label{table:lift}
\end{table}

\section{Conclusion and Future Work}
\label{sec:conclusion}
In this paper, we describe a non-invasive, machine learning based triaging tool for COVID-19. We collect and curate a large dataset of cough sounds with RT-PCR test results for thousands of individuals, and show with statistical evidence that our model can detect COVID-19 in the cough sounds from our dataset, even for patients that are entirely \textit{asymptomatic}. At current model performance, our tool can improve the testing capacity of a healthcare system by 43\%. Future work will involve incorporating other inputs from our dataset to the model, including breathing sounds, voice samples and symptoms.   
Our data collection is ongoing, and subsequent models will be trained on individuals beyond the subset in this study. We will also explore fast and computationally efficient inference, to enable COVID-19 testing on smartphones. This will enable large sections of the population to self-screen, support proactive testing and allow continuous monitoring.  

\section{Acknowledgments}

We are thankful to AWS for covering the entire cost of the GPUs used for this work. We also thank James Zhou (Stanford), and Peter Small) and Puneet Dewan (Global Health Labs)  for very helpful discussions, inputs, and evangelism. We are grateful to Ankit Baghel, Anoop Manjunath and Arda Sahiner from Stanford for helping with curating the cough pre-training dataset. 

We also want to thank the Governments of Bihar and Odisha, and the Municipal Corporation of Greater Mumbai for extending necessary approvals and facilitating activities for data collection in respective geographies. 

We are grateful to Ashfaq Bhat and his team at Norway India Partnership Initiative for supporting data collection and advocacy efforts in the state of Bihar and Odisha. Ravikant Singh and his team at Doctors for You for playing a critical role in initiating data collection, getting IRB approvals and managing field operations. Pankaj Bhardwaj and Suman Saurabh from Department of Community Medicine in All India Institute of Medical sciences, Jodhpur for leading the data collection efforts in the institute. 

We greatly appreciate the support of our lovely team members at Wadhwani AI. Nikhil Velpanur played a key role helping early data collection, supported by Akshita Bhanjdeo and Patanjali Pahwa. Puskar Pandey has been helping ensure continued data collection. Bhavin Vadera provided important support for data collection in additional sites. Vishal Agarwal helped build essential digital tools for data collection. Kalyani Shastry managed the entire logistics and coordinated supplies needed for field data collection at various study sites. 

And finally, we are humbled by the passion, hard work, and dedication of our numerous field staff. They have ensured strict adherence of the safety protocols through the data collection effort while maintaining high data quality. All this while working at the epicenters (hospitals and testing sites) of this global pandemic.

\bibliography{main}

\newpage
\appendix
\section{Data}
\begin{figure*}[!t]
  \centering
     \begin{subfigure}[b]{0.334\linewidth}
         \centering
         \includegraphics[width=\textwidth]{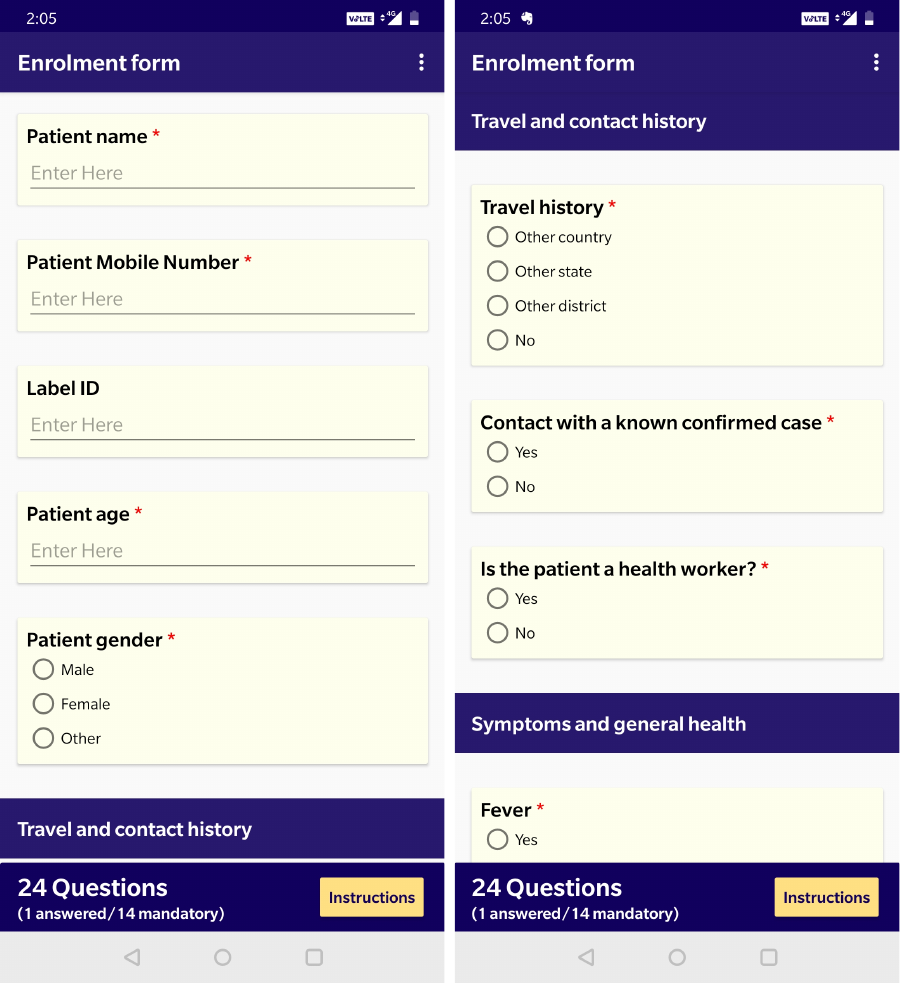}
         \caption{Enrollment app}
         \label{fig:enrollment-app}
     \end{subfigure}%
     \begin{subfigure}[b]{0.333\linewidth}
         \centering
         \includegraphics[width=\textwidth]{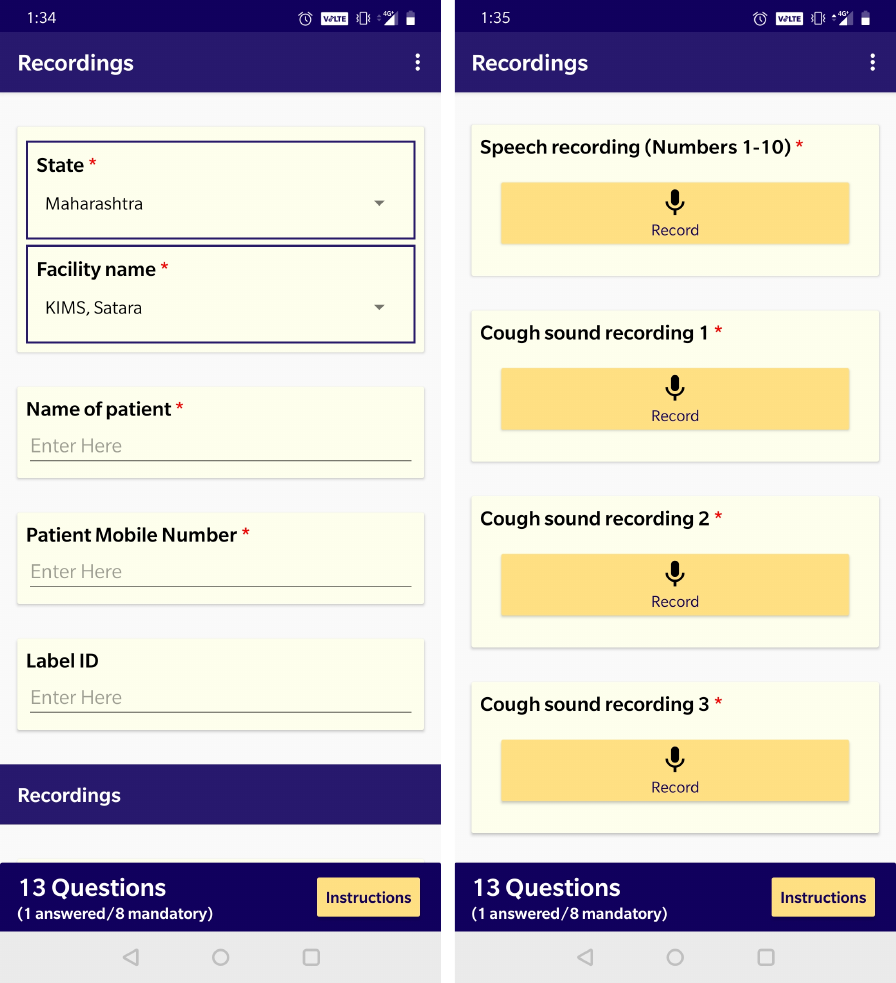}
         \caption{Audio Recording app}
         \label{fig:recordings-app}
     \end{subfigure}%
     \begin{subfigure}[b]{0.333\linewidth}
         \centering
         \includegraphics[width=\textwidth]{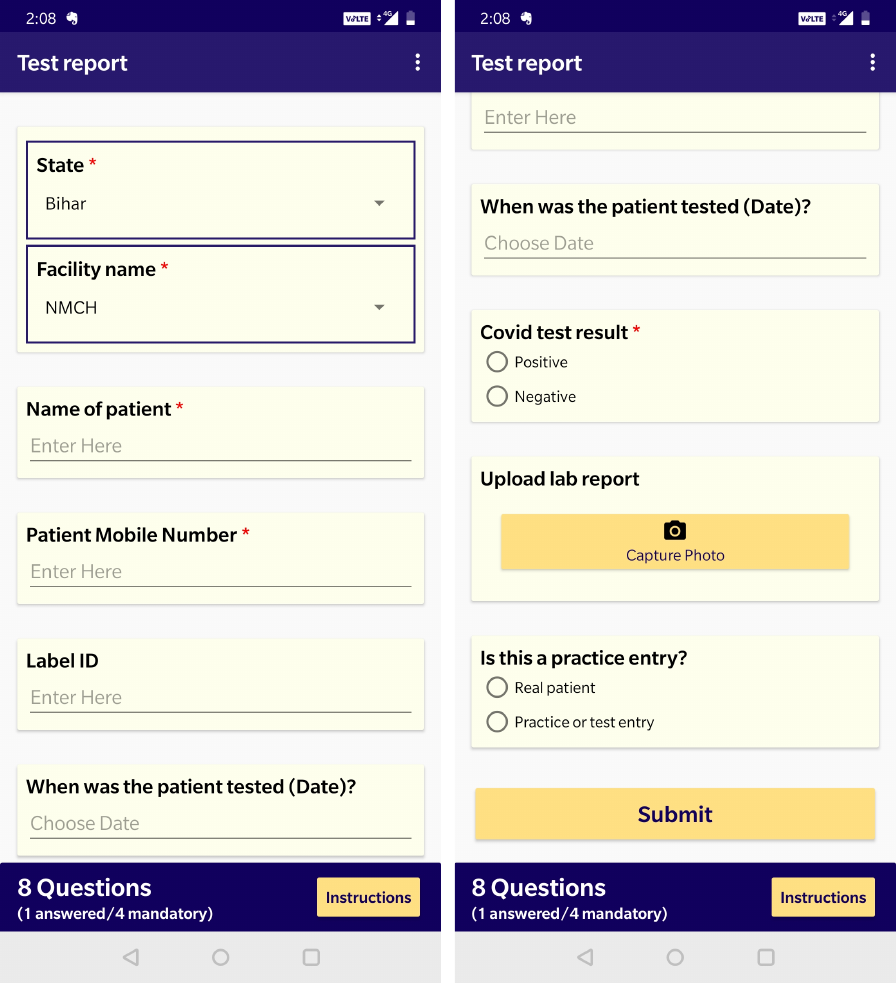}
         \caption{Test result app}
         \label{fig:testresult-app}
     \end{subfigure}
  \caption{\textit{Application Interfaces used in data collection}. Demographic, symptom and other health-related metadata are collected through (a) Our Enrollment app. Audio recordings are collected through (b) the Recording app and RT-PCR test results are uploaded for each patient using our (c) Test result app.}
  \label{fig:apps}
\end{figure*}
\subsection{Collection}
Our data collection pipeline consists of the following stages: (i) collection of individual specific meta data, (ii) recording of audio samples and finally (iii) obtaining the results of the COVID-19 RT-PCR test. We achieve this through three separate application interfaces as shown in Fig. \ref{fig:apps}. 
The details of data collected through these apps are enlisted below:

\begin{itemize}
    \item \textbf{Personal and Demographic information}: We collect the individual's name, mobile number, age, location (facility) and self-reported biological sex. 
    \item \textbf{Health-related information}: We collect the COVID-19 RT-PCR test result, body temperature and respiratory rate. We also note the presence of symptoms like fever, cough, shortness of breath and number of days since these symptoms first appear, and any measures undertaken specifically for cough relief. Finally, we also ask individuals if they have any co-morbidities. 
    \item \textbf{Additional metadata}: Additional data collected includes location (name of the facility, City and State), travel history of the individual, information about contact with confirmed COVID-19 cases, whether they are a health worker, and information about habits such as smoking, tobacco. 
\end{itemize}

\subsection{Preparation}

\subsubsection{Record linkage}: Since we use three different apps to collect data at different points in time, we need to link data across all three for a single individual. We achieve this through a semi-manual method that primarily uses fuzzy matching of each individual's name and phone number. Note that this process is non-trivial to automate since there are instances of wrongly entered texts, families sharing the same phone number etc. After the correspondence matching, we remove all identifiers from the dataset. 

\subsubsection{Manual validation}:  We manually validate each audio recording to check for errors in data collection. Specifically, for each cough, speech and breathing sample, we verify that the required sounds are actually present in the audio (e.g.\ cough sounds actually contain coughing). We only select the entries that pass this manual validation stage to create our usable dataset.


\subsection{Splits for Cross Validation}
Our dataset has a total of 1,039 individuals. We create 5 non-overlapping folds such that the validation set in each fold contains an equal number of positives and negatives from each facility. As noted in Sec. 6.6. of the paper, samples collected from different locations can have different label distributions. For example, testing facilities (F1, F3 and F4) tend to have predominantly COVID-19 negatives while isolation wards (F2) tends to contain COVID-19 positives (Fig. 2). In order to test that our model is not simply learning a facility classifier, we carefully curate the validation sets. We only consider data from the testing facilities F1 and F3 in the validation set. We do not test on facility F4 because of the small number of data samples obtained from this facility.

\section{Method}
\subsection{Reproducibility}
We set the seed as 42 for all packages that involve any randomness: PyTorch (\texttt{torch.cuda.manual\_seed\_all}, \texttt{torch.manual\_seed}), random (\texttt{random.seed}) and numpy (\texttt{np.random.seed}). This seed is set identically across all experiments. 

\subsection{Inference}
\subsubsection{File-level aggregation} 
Table \ref{tbl:aggregation} shows a comparison of various file-level aggregation methods (described in Sec. 4.4). Note that these numbers are without individual-level aggregation. Both median and mean perform equally well.

\subsubsection{Individual-level aggregation} Table. \ref{tbl:aggregation} shows the comparison of various individual-level aggregation methods (Sec. 4.4) with our ResNet-18 based model. We empirically find that max aggregation performs the best. 

\begin{table}[!h]
\begin{small}
  \centering
    \begin{tabular}{l|l|l}
    \hline
        \small{Method} & \small{File-level} & \small{Individual-level}    \\ \hline
        \texttt{min} & $0.62 \pm 0.05$ & $0.61 \pm 0.07$ \\ \hline
        \texttt{median} & $0.64 \pm 0.04$ & $0.65 \pm 0.07$  \\ \hline
        \texttt{mean}  & $0.64 \pm 0.05$ & $0.65 \pm 0.08$ \\ \hline
        \texttt{max}  & $0.62 \pm 0.03$ & $0.68 \pm 0.07$ \\ \hline
   \end{tabular}
  \caption{\textit{Comparison of aggregation methods.} For the file-level aggregation, \texttt{median} and \texttt{mean} over segment predictions seems to work equally well whereas for individual-level aggregation, \texttt{max} over probabilities over individual file-predictions works best.}
  \label{tbl:aggregation}
\end{small}
\end{table}

\subsubsection{Ensembling}
We tried two methods for ensembling:
\begin{itemize}
    \item \textbf{Ranking}: Ensembling uncalibrated models might lead to lower performance and since AUC doesn't require the predictions to be between 0 and 1, we rank the predictions instead of using the actual predicted probabilities. This gives us a minor performance lift from 0.680 to 0.686.
    \item \textbf{Stacked Ensemble}: As described in Sec. 5.5, we use XGBoost on top of the predictions from 3 models to improve the AUC from 0.68 to 0.72. The hyperparameters used for XGBoost are: \texttt{max\_depth=10, learning\_rate=0.1, n\_estimators=5000, scale\_pos\_weight=4000/pos\_ratio, min\_child\_weight=50, gamma=0.05, reg\_lambda=100}, where \texttt{pos\_ratio = 0.1} is the ratio of the number of positive samples to negative samples. The description of these parameters are given below:
    \begin{itemize}
        \item \texttt{max\_depth}: Maximum tree depth for base learners
        \item \texttt{learning\_rate}: Boosting learning rate
        \item \texttt{n\_estimators}: Number of gradient boosted trees. Equivalent to number of boosting rounds
        \item \texttt{scale\_pos\_weight}: Balancing of positive and negative weights
        \item \texttt{min\_child\_weight}: Minimum sum of instance weight (hessian) needed in a child
        \item \texttt{gamma}: Minimum loss reduction required to make a further partition on a leaf node of the tree
        \item \texttt{reg\_lambda}: L2 regularization term on weights
    \end{itemize}

    The descriptions for the full list of parameters and their default values can be found in the API documentation for XGBoost \footnote{https://xgboost.readthedocs.io/en/latest/python/python\_api.html\#module-xgboost.sklearn}.
\end{itemize}



\section{Use Case: COVID-19 triaging tool}
\subsubsection{Computation of lift from prevalence}
The utility of our method as a triaging tool for COVID-19 has been described in Sec. 7. Here, we show the detailed calculations  that we use to obtain the numbers for Table 4. 
The lift in testing capacity ($L$) is calculated as a function of the disease prevalence ($\rho$), the sensitivity ($S_{n}$), and the  specificity ($S_{p}$) of our model.

We use $n$ to denote the population size, and $TP$, $TN$, $FP$ and $FN$    to denote true positives, true negatives, false positives and false negatives respectively.
We propose a triaging mechanism wherein only individuals that are deemed positive from our model are sent for RT-PCR tests (Fig. 1, main paper). Hence all negatives from our model (which can be both true negatives $TN$ or false negatives $FN$) are not tested by RT-PCR. The number of false negatives from our model at the operating point we select (high sensitivity (90\%)) is extremely low.

Given we are not testing negatives, the effective increase (or lift) in testing capacity becomes 
\[
L = \frac{n}{n - (TN + FN)}
\]
It is trivial to show that 
\[
TN = S_{p}(1-\rho)n; \   FN = \rho n (1 - S_{n})
\]
Thus, we obtain the lift 
\[
L = \frac{1}{\left[ 1 - ((1 - \rho) S_{p}) + \rho  (1 - S_{n}) \right]}
\]

\end{document}